\documentclass[11pt]{article}
\usepackage{amsmath,amssymb,mathrsfs,color,epsfig,cite}
\usepackage{graphicx}
\usepackage{subfigure}
\usepackage{setspace}

\usepackage{amsfonts}

\newcommand{\hoch}[1]{$\, ^{#1}$}

\makeatletter
\@addtoreset{equation}{section}
\makeatother

\newcommand{\be}{\begin{equation}}
\newcommand{\ee}{\end{equation}}
\newcommand{\bea}{\setlength\arraycolsep{2pt} \begin{eqnarray}}
\newcommand{\eea}{\end{eqnarray}}
\newcommand{\nn}{\nonumber}

\def\ft#1#2{{\textstyle{\frac{\scriptstyle #1}{\scriptstyle #2} } }}
\def\fft#1#2{{\frac{#1}{#2}}}

\def\0{{\sst{(0)}}}
\def\1{{\sst{(1)}}}
\def\2{{\sst{(2)}}}
\def\3{{\sst{(3)}}}
\def\4{{\sst{(4)}}}
\def\5{{\sst{(5)}}}
\def\6{{\sst{(6)}}}
\def\7{{\sst{(7)}}}
\def\8{{\sst{(8)}}}
\def\sst#1{{\scriptscriptstyle #1}}

\def\ep{{\epsilon}}
\def\del{{\partial}}

\def\crampest{\medmuskip = 1mu plus 1mu minus 1mu}
\def\uncramp{\medmuskip = 4mu plus 2mu minus 4mu}

\def\im{{{\rm i\,}}}

\def\tal{{{\tilde\alpha}}}
\def\tkap{{{\tilde\kappa}}}
\def\tLam{{{\tilde\Lambda}}}
\def\tgam{{{\tilde\gamma}}}
\def\tnu{{{\tilde\nu}}}

\def\tx{{{\tilde x}}}

\def\tx{{{\tilde x}}}
\def\ty{{{\tilde y}}}
\def\cA{{{\cal A}}}

\def\scri{\mathscr{I}}

\begin{document}

\begin{flushright}
\hfill { UPR-1320-T \ \ \ CERN-TH-2022-102\ \ \ MI-HET-779
}\\
\end{flushright}

\begin{center}
{\large {\bf Mode Stability For Massless Scalars In Five-Dimensional 
Black Hole Backgrounds
 }}

\vspace{15pt}
{\large M. Cveti\v c$^{1,2,3,4}$, 
              C.N. Pope$^{5,6}$, B.F. Whiting$^{7}$ and Haoyu Zhang$^{5}$}

\vspace{15pt}

{\hoch{1}}{\it Department of Physics and Astronomy,
University of Pennsylvania, \\
Philadelphia, PA 19104, USA}

{\hoch{2}}{\it Department of Mathematics, University of Pennsylvania, Philadelphia, PA 19104, USA}

{\hoch{3}}{\it Center for Applied Mathematics and Theoretical Physics,\\
University of Maribor, SI2000 Maribor, Slovenia}

{\hoch{4}}{\it CERN Theory Department, CH-1211 Geneva, Switzerland}


\hoch{5}{\it George P. \& Cynthia Woods Mitchell  Institute
for Fundamental Physics and Astronomy,\\
Texas A\&M University, College Station, TX 77843, USA}

\hoch{6}{\it DAMTP, Centre for Mathematical Sciences,
 Cambridge University,\\  Wilberforce Road, Cambridge CB3 OWA, UK}

\hoch{7}{\it Department of Physics, P.O. Box 118440, University of Florida,\\
Gainsville, FL 32611-8440, USA}

\vspace{10pt}



\end{center}




\begin{abstract}

The mode stability of the Kerr black hole in four dimensions was demonstrated
by Whiting in 1989, by separating the Teukolsky equation that describes
gravitational perturbations and then transforming the radial and angular equations in such a way that the problem can be reformulated as a wave 
equation in an auxiliary spacetime in which the proof of stability is
greatly simplified, owing to the absence of an ergoregion.  As a preliminary 
step towards extending these ideas to higher-dimensional black holes,
we study the mode stability of the massless scalar wave equation in
the five-dimensional black hole solutions of Einstein gravity and 
supergravity. We show how the wave equation can again be mapped into one
in an auxiliary spacetime in which there is no ergoregion, allowing us
to give a proof of the mode stability of the solutions of the 
scalar wave equation.

\end{abstract}

{\small 
cvetic@physics.upenn.edu, pope@physics.tamu.edu, bernard@phys.ufl.edu,
zhanghaoyu@tamu.edu.}
\pagebreak

\tableofcontents
\addtocontents{toc}{\protect\setcounter{tocdepth}{2}}

\section{Introduction}

  Establishing results on the stability of black hole solutions has been a
major activity in the general relativity community for many years.  This
is especially subtle, and also important, in the case of stationary
rotating black holes.  Many
different approaches have been followed, but one of the most fruitful
has involved finding an integral transformation that maps the difficult
analysis in the original variables into a considerably simpler analysis
in the transformed variables.  The idea was first developed in 
\cite{whiting}, where it was employed to establish the mode stability of
the Kerr black hole \cite{whiting}.  The technique has been developed 
further in recent years, and found application in studies such as the global
stability of black holes (see, for example, \cite{dafermosg,shlapcost}), 
the stability of extremal black holes \cite{rita}, and 
mode stability on the real frequency axis \cite{Shlapentokh-Rothman:2013hza, larsetal}.  

   Establishing the mode stability of the Kerr black hole involved 
studying the properties of the mode functions in the separation of 
variables for the Teukolsky equation, which provides a gauge-invariant 
description of the perturbations around the Kerr background.  In
\cite{whiting} the  
generalised Teukolsky equation with a spin parameter $s$ was studied, with
$s=\pm2$ corresponding to the actual case of interest in which the 
equation describes the gravitational perturbations themselves.  The
case $s=0$ corresponds to the massless scalar wave equation (the massless
Klein-Gordon equation), 
while the $s=\pm 1$ case governs gauge-invariant components of the Maxwell field.  

   The techniques for analysing the Teukolsky
equation for spin $s$ that were developed in \cite{whiting} were broadly
similar for all $s$, and in fact the essential features associated with
the stability of the solutions could already be seen in the $s=0$ case.
This is a useful observation because if one looks at more complicated
situations than black holes in pure Einstein gravity, such as black holes
in Einstein-Maxwell theory or in supergravity, the analogue of the
Teukolsky analysis has not been implemented.  In the absence of a
gauge-invariant treatment of the perturbations of the black holes in
these more complicated situations, one can at least study the 
analogue of $s=0$ Teukolsky equation, that is, the massless scalar
wave equation in the black hole background.  One may hope that this can
provide a ``proxy'' for the full analysis, and that establishing stability
results for the solutions of the massless wave equation may be 
indicative of what one might find in a more elaborate and complete
perturbative analysis.  This approach was adopted in \cite{cvgipowh},
where the techniques of \cite{whiting} were applied to the study of
the mode stability for solutions of the massless scalar wave equation in
the background of a class of four-dimensional supergravity black holes
carrying four independent electric charges \cite{Cvetic:1996kv,Chong:2004na}.

  In this paper, we extend some of the four-dimensional techniques for
mode-stability analysis that were developed in \cite{whiting} to the
case of five dimensions.  Even in the pure Einstein case, the
analogue of the four-dimensional Teukolsky analysis is unknown.  The
stumbling block is that the four-dimensional analysis depended heavily
upon the use of the Newman-Penrose formalism, and no particularly useful
extension of this to five or higher dimensions has been 
constructed.\footnote{However, see \cite{Gomez-Lobo:2009riu}.} 
Thus for now, our approach will mirror the one that can be followed
for more complicated theories in four dimensions, namely, we shall focus
attention on establishing stability results for solutions of the
five-dimensional massless scalar wave equation.  This already allows us
to develop a rather non-trivial generalisation of the techniques that
were employed in four dimensions, and it reveals ways in which the 
integral transformation that allows us to establish mode-stability results
is substantially different from the one in four dimensions.

  We shall describe in the subsequent sections how one can establish 
mode stability results for solutions of the massless wave equation in
the background of the five-dimensional Myers-Perry rotating black hole \cite{myper},
and also in the background of general 3-charge rotating black holes
in the five-dimensional STU supergravity theory \cite{cvetyoum5}.

\section{Massless Scalar 
Wave Equation in Five-Dimensional Black Hole Background}

\subsection{Five-dimensional Myers-Perry black hole}

  The natural generalisation of the four-dimensional
rotating Kerr \cite{kerr} 
black hole to higher spacetime dimensions is provided by
the Myers-Perry black hole solutions \cite{myper}.  These vacuum solutions of
the $D$-dimensional Einstein equations are characterised by their mass $M$
and by $[(D-1)/2]$ independent angular momenta, reflecting the fact that
independent rotations can occur in each orthogonal spatial 2-plane.  In this
paper we shall be concerned specifically with the example of the 
five-dimensional rotating black hole.  Its metric is given by \cite{myper}
\bea
ds^2 &=& -\fft{\Delta}{\rho^2}\, (dt - a \sin^2\theta\, d\phi -b \cos^2\theta\,
d\psi)^2  +
  \fft{\rho^2}{\Delta}\, dr^2 + \rho^2\, d\theta^2\nn\\
&&
+ \fft{\sin^2\theta}{\rho^2}\, [a dt -(r^2+a^2)\, d\phi]^2
 +\fft{\cos^2\theta}{\rho^2}\, [b dt -(r^2+b^2)\, d\psi]^2 \nn\\
&& 
+\fft1{r^2\, \rho^2}\, [a b dt -b (r^2+a^2) \sin^2\theta d\phi -
    a(r^2+b^2)\cos^2\theta\, d\psi]^2\label{d5mp}
\eea
where
\bea
\Delta= \fft{(r^2+a^2)(r^2+b^2)}{r^2} -2M\,,\qquad
\rho^2= r^2+ a^2\, \cos^2\theta + b^2\, \sin^2\theta\,.
\eea
Here $a$ and $b$ are the two independent rotation parameters, with $\phi$ and
$\psi$ being the two associated azimuthal angles (each with period $2\pi$).
The latitude coordinate $\theta$ ranges over $0\le\theta\le\ft12\pi$.

\subsection{Massless scalar wave equation}

  Our focus will be the investigation of solutions of the massless scalar
wave equation $\square\Psi=0$ in the Myers-Perry background, with the goal
of establishing that modes with time dependence $e^{-\im\omega t}$ 
that are ingoing on the future horizon and outgoing at future null infinity
cannot have a frequency $\omega$ with a positive imaginary part. In other
words, we seek to show that there cannot exist spatially regular modes
that would give rise to instabilities growing exponentially in time.   

  One can define new radial and latitude coordinates $\tx$ and $y$ by 
writing
\be
\tx=r^2\,,\qquad y=\cos^2\theta\,.
\ee
Using these, we have 
\be
\sqrt{-g} = \fft{\tx + a^2\, y+ b^2\, (1-y)}{4}\,,
\ee
and defining the quantity $G^{\mu\nu}\equiv \sqrt{-g}\, g^{\mu\nu}$, we have
\bea
G^{00}&=& -\fft{(\tx+2M + b^2)}{4} - \fft{M^2\, \tx}{D} 
+ \fft{(b^2-a^2)y}{4}\,,
 \quad G^{01}=G^{02}=0\,,\quad G^{03}= -\fft{a M (\tx+b^2)}{2D}\,,\nn\\
G^{04} &=& -\fft{b M(\tx+a^2)}{2D}\,,\quad
G^{11}= \tx^2+ (a^2+b^2-2M)\tx + a^2 b^2\,,\quad
G^{12}= G^{13}=G^{14}=0\,,\nn\\
G^{22}&=& y(1-y)\,,\quad G^{23}=G^{24}=0\,,\nn\\
G^{33}&=& \fft{(b^2-a^2) \tx + b^2\, (b^2-a^2 -2M)}{4D} +\fft1{4(1-y)}\,,\quad
G^{34} = -\fft{a b M}{2D}\,,\nn\\
G^{44} &=& \fft{(a^2-b^2) \tx + a^2\, (a^2-b^2 -2M)}{4 D} + \fft1{4y}\,,
\label{sqinv}
\eea
where
\be
D= (\tx+a^2)(\tx+b^2) -2M \tx\,. \label{DD}
\ee
It
can be shown \cite{Cvetic:1997uw}  that the massless scalar wave equation 
$\square\Psi=0$, which may be written as $\del_\mu(G^{\mu\nu}\, \del_\nu
\Psi)=0$, is separable. 

  We can obtain the separated $\tx$ and $y$ equations in Schr{\" o}dinger 
form by writing \footnote{For the separated form of the equations before  casting them in  the Schr{\" odinger} form, see \cite{Cvetic:1997uw}. There the separation was already performed for the 5d STU black holes \cite{cvetyoum5}.}
\be
\Psi= e^{-\im\omega t + \im m \phi + \im n\psi}\, R(\tx)\, S(y)
\ee
and then defining
\be
R(\tx)= \fft{X(\tx)}{\sqrt{D}} \,,\qquad S(y) =\fft{Y(y)}{\sqrt{y(1-y)}}\,.
\ee
This gives the separated equations
\be
\fft{X''}{X}  + U_\tx - \fft{\sigma}{D}=0\,,\qquad 
\fft{Y''}{Y} + U_y + \fft{\sigma}{y(1-y)}=0\,.\label{XYeqn}
\ee

  It is convenient now to make a further change of the radial variable.
 Defining $\ep_+$ and $\ep_-$ by
\bea
2M -(a+b)^2 = 2M \ep_+^2\,,\qquad 2M -(a-b)^2= 2M \ep_-^2\,,\label{eppepm}
\eea
a further coordinate transformation from $\tx$ to $x$, given by
\bea
\tx =  2M \ep_+\, \ep_-\, x + \ft12 M (\ep_+-\ep_-)^2\,,\label{xdefine}
\eea
implies that $D$, defined in eqn (\ref{DD}), becomes
\bea
D= 4 M^2\, \ep_+^2\, \ep_-^2\, x(x-1)\,.
\eea
The outer horizon $r=r_+$ 
is located at $x=1$, and the inner horizon $r=r_-$ at $x=0$.
The region where $r$ goes to infinity corresponds to $x$ goes to infinity.

 In terms of the variable $x$, 
we can now write the radial equation in the form
\bea
\fft{X''}{X} +\fft{\kappa+\Lambda}{x} + \fft{\kappa-\Lambda}{x-1} + 
\fft{\ft14 -\beta^2}{x^2} + \fft{\ft14-\gamma^2}{(x-1)^2}=0\,,\label{Xeqn}
\eea
finding 
\bea
\kappa &=&\ft14 M \ep_+\,\ep_-\, \omega^2\,,\nn\\
\beta &=& \fft{\im}{4\sqrt{2M}}\, 
 \Big[\fft{2M\omega -(a+b)(m+n)}{\ep_+} -
     \fft{2M\omega -(a-b)(m-n)}{\ep_-}\Big]\,,\nn\\
\gamma &=& \fft{\im}{4\sqrt{2M}}\, 
 \Big[\fft{2M\omega -(a+b)(m+n)}{\ep_+} +
     \fft{2M\omega -(a-b)(m-n)}{\ep_-}\Big]\,,\label{greek1}\\
\Lambda &=& \sigma+ \fft12 - \fft34 M \omega^2 + \ft18 (a^2-b^2)\omega^2 +
 \fft{[2M\omega-(a+b)(m+n)]^2}{16M\ep_+^2} +
  \fft{[2M\omega-(a-b)(m-n)]^2}{16M\ep_-^2}\,.\nn
\eea

In the angular direction, the equation for $Y$ in (\ref{XYeqn}) is of the
form
\bea
\fft{Y''}{Y} + \fft{\hat\kappa+\hat\Lambda}{y} + 
   \fft{\hat\kappa-\hat\Lambda}{y-1} +
\fft{\ft14 -\hat\beta^2}{y^2} + \fft{\ft14-\hat\gamma^2}{(y-1)^2}=0\,,
\label{Yeqn}
\eea
with
\bea
\hat\kappa &=& -\fft18 (a^2-b^2)\omega^2\,,\qquad \hat\beta=\fft{n}{2}\,,\qquad
\hat\gamma= \fft{m}{2}\nn\\
\hat\Lambda &=& \fft12 +\sigma - \fft{(m^2+n^2)}{4} + 
  \fft{(a^2-b^2)\omega^2}{8}\,.\label{hattedgreeks}
\eea

\subsection{Transforming the radial equation}

  We start with the radial equation given by eqn (\ref{Xeqn}), 
with the constants given in (\ref{greek1}).  Next, introduce a new 
function $f(x)$, related to $X(x)$ by
\bea
X(x) = x^{\fft12  - \ep_1\, \beta}\, (x-1)^{\fft12-\ep_2\,\gamma}\, f(x)\,,
\eea
where $\ep_1^2=\ep_2^2=1$.
The function $f$ therefore satisfies
\bea
[x(x-1)\, \del^2_{x} + (bx + c) \,\del_x + dx + e] f(x)=0\,,\label{feqn}
\eea
with
\bea
b&=& 2(1-\ep_1\,\beta-\ep_2\,\gamma)\,,\qquad c= -1 + 2\ep_1\,\beta\,,\nn\\
d &=& 2\kappa\,,\qquad e= \ft12 -\kappa -\ep_1\, \beta -\ep_2\,\gamma + 
   2\ep_1\,\ep_2\,\beta\,\gamma -\Lambda\,.\label{bcde}
\eea

Next, we make an integral transform to a new radial variable $h(z)$, by 
defining
\bea
h(z) = e^{-\tal\, z}\, e^{\tnu/z}\, z^{1+\tgam}\, 
   \int_1^\infty e^{2\tal\,x\,z}\, f(x)\, dx\label{hdef}
\eea
Now, multiplying eqn (\ref{feqn}) by $e^{2\tal\, x\, z}$ and integrating
gives, after integration by parts,
\bea
0= B + \int_1^\infty [4\tal^2 \, x(x-1)\, z^2 +
        2\tal\, (4x-bx -2-c)z + dx + 2-b+e]\, f(x)\, dx\,,\label{tran1}
\eea
where $B$ is the boundary term
\bea
B= \Big[e^{2\tal\, x\, z} x(x-1)\,\del_x f - \del_x(e^{2\tal\, x\,z}\, 
x(x-1)) \, f+ e^{2\tal\,x\,z} (bx+c) \, f\Big]_1^\infty\,.\label{boundaryterm}
\eea
Note that eqn (\ref{tran1}) can be written as
\bea
0=B +\int_1^\infty f(x) {\cal O}_z(e^{2\tal\,x\,z})\, dx\,,
\eea
where
\bea
 {\cal O}_z = z^2\, \del^2_{z} + \left(4z-2\tal \, z^2 -b z + \fft{d}{2\tal}\right)\,\del_z
+ 2-2\tal\, c\, z-4\tal\, z  -b+e\,.
\eea

It can then be seen that provided the boundary term $B$ vanishes, a question to 
which we shall return later, then $h(z)$ defined in eqn (\ref{hdef}) satisfies
${\cal O}_z\Big(e^{\tal\, z}\, e^{-\tnu/z}\, z^{-1-\tgam}\, h(z)\Big)=0$,
and thus
\bea
\Big(\del^2_{z} -\tal^2 + \fft{2\tal\,\tkap}{z} + \fft{\tLam}{z^2} +
  \fft{2\tgam\, \tnu}{z^3} - \fft{\tnu^2}{z^4}\Big) h(z)=0\,,
\label{heqn}
\eea
where the constants are chosen so that
\bea
\tkap&=& -\ep_1\, \beta + \ep_2\,\gamma\,,\qquad
\tLam= \ft12 - \beta^2 -\gamma^2 -\Lambda\,,\nn\\
\tgam&=& \ep_1\, \beta +\ep_2\, \gamma\,,\qquad
\tnu= -\fft{\kappa}{2\tal}\,.\label{tgreeks}
\eea

  The constant $\tal$ is arbitrary at this stage (it really just sets the
scale of the new radial variable $z$ that is introduced in eqn (\ref{hdef})).
We shall find it convenient to define it to be
\bea
\tal = \fft{\im \sqrt{M}\, \ep_-\, \omega}{2\sqrt2}\,.\label{taldef}
\eea

\subsection{Behaviour of unstable modes}

  If they existed, unstable modes would be solutions than were purely 
outgoing at $\scri$ and purely ingoing at the horizon 
${\cal H}$, that is, asymptotically, they would have support only at $\scri^+$ 
and ${\cal H}^+$.  We are interested in establishing the non-existence
of unstable modes arising 
from frequencies in the upper half of the complex $\omega$-plane, that is for 
$\omega=\omega_0+i\omega_1$, with $\omega_1>0$, since these would grow
exponentially in the future.  Thus, near $\scri^+$, we must have:
\be
\Psi\sim\exp{-i\omega(t-r_*)}\,,\qquad \hbox{as}\quad 
 \{t,r_*\}\rightarrow\infty\,,
\ee
and, near ${\cal H}^+$, we must have:
\be
\Psi\sim\exp{-i\omega(t+r_*)}\,,\qquad \hbox{as}\quad
\{-t,r_*\}\rightarrow-\infty\,.
\ee
 From the definition of the $r_*$ coordinate in eqn (\ref{r*def}), it can be
seen that in terms of the radial coordinate $x$ introduced in eqn 
(\ref{xdefine}), we shall have asymptotically
\bea
\hbox{Near}\quad \scri:&&\qquad r_*= \sqrt{2M \ep_+\,\ep_-}\, x^{\fft12} +
  {\cal O}\big(x^{-1/2}\big)\,,\label{r*inf}\\
\hbox{Near}\quad {\cal H} :&&\qquad r_* =
   \fft{\sqrt{M}\, (\ep_++\ep_-)}{2\sqrt2\, \ep_+\,\ep_-}\, \log(x-1) +
{\cal O}\big(x-1\big)\,.\label{r*hor}
\eea

  From the radial equation (\ref{Xeqn}), it can be seen that near
the horizon $x=1$, the solutions will have the form
\bea
X(x) = (x-1)^{\fft12 \pm\gamma}\, F^{(1)}_\pm(x-1)\,,\label{hormodes}
\eea
where the functions $F^{(1)}_\pm(x-1)$ are analytic around $x=1$, 
with $F^{(1)}_\pm(0)$ a non-zero constant.   As noted above, an
unstable mode would have $\omega$ dependence of the form 
$e^{-\im \omega r_*}$ near {\cal H}, and hence from (\ref{r*hor}) and (\ref{greek1})
it would correspond to the minus-sign choice in eqn (\ref{hormodes}).
Thus for an unstable mode $X_{um}$, we must have the near-horizon behaviour
\bea
\hbox{Near}\quad {\cal H}: \qquad X_{um}(x) = 
(x-1)^{\fft12 -\gamma}\, F^{(1)}_-(x-1)\,.\label{umhor}
\eea

   In the asymptotic region near $x=\infty$, it can be seen from the radial
equation (\ref{Xeqn}) that the asymptotic form of the solutions will be
\bea
X(x) = x^{\fft14}\, e^{\pm 2\im\sqrt{2\kappa}\, \sqrt{x}}\, 
F^{(\infty)}_\pm\Big(\fft1{\sqrt{x}}\Big)\,,\label{infmodes}
\eea
with the functions 
$F^{(\infty)}_\pm\Big(\fft1{\sqrt{x}}\Big)$ being asymptotic series
with $F^{(\infty)}_\pm(0)$ a non-zero constant.  Now, as we saw, 
an unstable mode would
correspond to a having an $\omega$ dependence of the form $e^{\im \omega r_*}$
near infinity, and so 
from (\ref{r*inf}) and (\ref{greek1}) it would
correspond to the plus-sign choice in eqn (\ref{infmodes}).  Thus, for
an unstable mode $X_{um}(x)$, we must have the asymptotic behaviour
\bea
\hbox{Near}\quad \scri:\qquad 
X_{um}(x) = x^{\fft14}\, e^{2\im\sqrt{2\kappa}\, \sqrt{x}}\,
F^{(\infty)}_+\Big(\fft1{\sqrt{x}}\Big)\,.\label{uminf}
\eea
Before going on, we point out that infinitely many  
modes with these analytic properties do exist with $\omega_1$, the imaginary
part of $\omega$, being $<0$.  These modes decay exponentially in time, 
have been extensively studied \cite{Kokkotas:1999bd}, and 
are known as the quasi-normal modes which, taken together, uniquely 
characterise a black hole spacetime to which they correspond. 

   We turn now to the behaviour of the transformed radial function
$h(z)$, defined by eqn (\ref{hdef}).  In particular, we shall be 
concerned with the behaviour in the coordinate range $0\le z\le \infty$.  

   Near $z=0$, it can be seen from eqn (\ref{heqn}) that the leading-order
behaviour of the solutions will be
\bea
h(z) = e^{\pm\tnu/z}\, z^{1\pm\tgam}\, G^{(0)}_\pm (z)\,,\label{hzero}
\eea
where $G^{(0)}_\pm (z)$ are analytic functions with $G^{(0)}_\pm (0)$
being non-zero constants.  It is evident from (\ref{hdef}) that at 
$z=0$ the 
integrand for an unstable mode is just
an analytic function of $x$, and so the leading-order
behaviour of $h(z)$ near $z=0$ will be given by the pre-factor
functions $e^{\tnu/z}\, z^{1+\tgam}$. In other words, the unstable mode 
corresponds to the plus-sign choice in eqn (\ref{hzero}):
\bea
\hbox{Near}\quad z=0:\qquad
  h_{um}(z) = e^{\tnu/z}\, z^{1 +\tgam}\, G^{(0)}_+ (z)\,,\label{hum}
\eea
As a check, we see from eqns (\ref{greek1}), (\ref{tgreeks}) 
and (\ref{taldef}) that
\bea
\tnu = \fft{\im \sqrt{M}\, \ep_+\,\, \omega}{2\sqrt2}\,,
\eea
whose real part is negative when $\omega$ has a positive imaginary part,
thus implying that $h_{um}(z)$ in eqn (\ref{hum}) is finite, and goes to
zero, as $z$ goes to zero.  (This justifies the sign choice in the 
definition of $\tal$ in eqn (\ref{taldef}).)

    It can be seen from eqn (\ref{heqn}) that near $z=\infty$, the function
$h(z)$ has the behaviour
\bea 
h(z) = e^{\pm\tal z}\, z^{\mp\tkap}\, G^{(\infty)}_\pm\Big(\fft1z\Big)\,,
\label{hinf}
\eea
where the functions $G^{(\infty)}_\pm\Big(\fft1z\Big)$ are asymptotic
in $z^{-1}$ with $G^{(\infty)}_\pm(0)$ being non-vanishing constants.  
In the expression (\ref{hdef}) the leading behaviour of $h(z)$
near $z=\infty$ is governed by the behaviour of $f(x)$ near $x=1$.  
Using the previously-determined behaviour of $X(x)$, and hence $f(x)$,
for an unstable mode we see that the integrand in eqn (\ref{hdef}) 
has the behaviour
\bea
\int_1^\infty e^{2\tilde{\alpha}x z} f(x) dx&=
 &e^{2\tilde{\alpha}z}\int_1^\infty e^{2\tilde{\alpha}(x-1)z}\,
  (x-1)^{(\epsilon_2-1)\gamma}\Big(1+{\cal O}(x-1)\Big) dx.
\eea
Substituting $v=-2\tilde{\alpha}(x-1)z$, we then have:
\bea
\int_1^\infty e^{2\tilde{\alpha} xz} f(x) dx&\propto&
 e^{2\tilde{\alpha}z}\,z^{-1+\gamma(1-\epsilon_2)}
 \int_0^\infty e^{-v}v^{(\epsilon_2-1)\gamma}
\Big(1+{\cal O}\left(\frac{v}{z}\right)\Big) dv.
\eea
Combining with the pre-factor we then see, provided 
\bea
\epsilon_2=-1\,,\label{ep2}
\eea
that:
\bea
h(z)&\propto&e^{+\tilde{\alpha} z}\,y^{\tilde{\gamma}+
  (1-\epsilon_2)\gamma}(1+{\cal O}\left(\frac{1}{z}\right)\Big)\nn\\
&\sim&e^{\tilde{\alpha} z}\,z^{-\tilde{\kappa}}\,.
\eea
Thus for an unstable mode, the plus-sign choice in eqn (\ref{hinf}) is
selected:
\bea
\hbox{Near}\quad z=\infty:\qquad 
h_{um}(z) = e^{\tal z}\, z^{-\tkap}\, G^{(\infty)}_+\Big(\fft1z\Big)\,.
\label{huminf}
\eea
As a check, we note that if $\omega$ has a positive imaginary part, 
$\tal$, given in eqn (\ref{taldef}), will have a negative real part,
and so $h_{um}(z)$ will be finite as $z$ goes to infinity.  (This
motivated the sign choice in the definition of $\tal$ in eqn (\ref{taldef}).)

    It is now straightforward to check, using the asymptotic 
properties of the radial functions for unstable modes established in this
section, that the boundary term given in eqn (\ref{boundaryterm})
will vanish for any unstable mode.  Thus we have established that there
is a one-to-one mapping between exponentially unstable modes in the original 
untransformed radial function $X(x)$ and exponentially unstable modes in the
radial function $h(z)$ obtained by means of the integral transform
(\ref{hdef}).  The final steps in the proof of the mode stability will
be presented in the next section; this will entail establishing for
the transformed radial equation that there cannot exist any exponentially unstable modes.

\section{Modes in the Transformed Spacetime}

\subsection{Combining angular and transformed radial equations}

The constant $\sigma$ that was introduced in the original process of 
separating variables is present in the transformed radial equation (\ref{heqn})
through the quantity $\Lambda$ (see eqns (\ref{greek1}) and 
(\ref{tgreeks})) and in the angular equation (\ref{Yeqn}) through
the quantity $\hat\Lambda$ (see eqn (\ref{hattedgreeks})).  It follows
therefore that if we form the combination 
\bea
\fft{z^2}{h(z)}\, \del^2_{z}\, h(z) + \fft{y(1-y)}{Y(y)}\, \del^2_{y}\, Y(y)
\label{hYcomb}
\eea
and make use of eqns (\ref{heqn}) and (\ref{Yeqn}) 
then we shall obtain an equation in which all the $\sigma$ dependence has
cancelled.  This equation can in fact be interpreted as the result
of performing a separation of variables in which we write
\bea
\Psi(t,z,y,\phi,\psi) = h(z)\, Y(y)\, e^{-\im\omega t}\, e^{\im m \phi}\,
    e^{\im n \psi}\,.
\eea

   We postpone writing the full `unseparated'' equation for now, and
just focus on the terms proportional to $\omega^2$.  (That is, the
$-\del_{tt}$ terms in the full five-dimensional wave equation.)  Together
with the terms involving the radial and angular derivatives these
are 
\bea
&& z^2\, \del^2_{z} + y(1-y)\, \del^2_{y} + 
 \fft{\ep_+^2\,M\,\omega^2}{8 z^2} + 
\fft{\ep_-^2\, M\, z^2\, \omega^2}{8} +  
\fft{[-\ep_1\, (\ep_- -\ep_+) -\ep_2\,
   (\ep_- +\ep_+)]\,M\,\omega^2}{4\, \ep_-\, z} \nn\\
&&+ \fft{[\ep_1\, (\ep_- -\ep_+) - \ep_2\, (\ep_- + \ep_+)]
   \, M\, z\, \omega^2}{4\, \ep_+} 
   + \ft34 M\, \omega^2 - \ft18 (a^2-b^2)\, (1-2y)\, \omega^2 +
  \hbox{rest} \,,\label{omsq}
\eea
Since 
\bea
\ep_+ = \sqrt{1 -\fft{(a+b)^2}{2M}}\,,\qquad 
 \ep_- = \sqrt{1 -\fft{(a-b)^2}{2M}}\,,\label{eppepm2}
\eea
(see eqns (\ref{eppepm})), and we always assume $a$ and $b$ are non-negative,
it follows that $\ep_-\ge \ep_+$.  Consequently, all the $\omega^2$ terms in
eqn (\ref{omsq}) will be non-negative, provided 
that we choose the sign of $\epsilon_2$ to be
\bea
\ep_2=-1\,.\label{ep22}
\eea
Note that the necessity for this choice of sign was already seen in the
previous section, in eqn (\ref{ep2}).  
The choice of sign for $\ep_1$ is undetermined by these considerations. We
shall, for definiteness, make the choice
\bea
\ep_1=-1
\eea
For the remaining $\omega^2$ terms, namely with coefficient
\bea
\ft34 M - \ft18 (a^2-b^2)(1-2y)\,,\label{remterms}
\eea
we note that the range of the angular coordinate is $0\le y\le1$, and
so $-1\le 1-2y \le 1$. We also have that
\bea
(a+b)^2 \le 2M\,,
\eea
(see eqn (\ref{eppepm2})), and so we have $a^2-b^2= (a+b)(a-b)\le (a+b)^2
\le 2M$.  Thus the remaining terms (\ref{remterms}) contributing to
$\omega^2$ terms in (\ref{omsq}) are always positive.

  In summary, we have seen that the overall coefficient of $\omega^2$ in
eqn (\ref{omsq}) is always positive, implying that in the transformed
metric $\tilde g_{\mu\nu}$ obtained by the process of
unseparating variables, $\fft{\del}{\del t}$ is always timelike outside the
horizon.

 We now present the complete result for the combination of the radial and
angular equations.  With $\ep_1=\ep_2=-1$ as discussed above we obtain
\bea
&&\Big\{z^2\, \del^2_{z} + y(1-y)\,\del^2_{y} -
\Big(\fft{(a+b)}{4 z} + 
    \fft{(a-b)\, z}{4}\Big)\, m\omega
-\Big(\fft{(a+b)}{4 z} - 
   \fft{(a-b)\, z}{4}\Big)\, n\omega\nn\\
&& +   \Big( \fft{\ep_+^2\,M}{8z^2} + \fft{\ep_-^2\,M\,  z^2}{8} + 
  \fft{M}{2 z} + \fft{M\,z}{2}  +
  \ft34 M - \ft18(a^2-b^2)(1-2y)\Big)\,\omega^2\nn\\
&& -\fft{m^2}{4(1-y)} -\fft{n^2}{4 y} + \fft1{4y(1-y)}\Big\} \Psi=0\,.
\label{comb}
\eea

Using the replacements
\bea
\omega\longrightarrow \im\, \del_t\,,\qquad
m\longrightarrow -\im\, \del_\phi\,,\qquad
n\longrightarrow -\im\, \del_\psi
\eea
we can read off 
the components $\tilde g^{\mu\nu}$ of an inverse metric in a
transformed spacetime, such that eqn (\ref{comb}) can be written as
\bea
\tilde g^{\mu\nu}\,\del_\mu\, \del_\nu \Psi + \fft1{4y(1-y)}\,\Psi=0\,,
\label{top}
\eea
with
\bea
\tilde g^{zz}&=& z^2\,,\qquad \tilde g^{yy} = y(1-y)\,,\quad 
\tilde g^{\phi\phi} = \fft1{4(1-y)}\,,\quad
\tilde g^{\psi\psi}= \fft1{4y}\,,\nn\\
\tilde g^{t\phi} &=& -\fft{(a+b)}{8 z} -
    \fft{(a-b)\, z}{8}\,,\quad
\tilde g^{t\psi}= -\fft{(a+b)}{8z} +
    \fft{(a-b)\, z}{8}\,,\nn\\
\tilde g^{tt}&=& - \Big( \fft{\ep_+^2\, M}{8z^2} + 
\fft{\ep_-^2\, M\,  z^2}{8} +
  \fft{M}{2z} + \fft{M\,z}{2}  +
  \ft34 M - \ft18(a^2-b^2)(1-2y)\Big)\,.\label{tildemet}
\eea

   We may find a suitable conformal factor $\Omega^2$ and a redefined
wave function $\Phi$ such that the
D'Alembertian of $\Phi$ in a rescaled metric $\hat g_{\mu\nu}= \Omega^2\, 
\tilde g_{\mu\nu}$ gives rise to eqn (\ref{top}).   We define
\bea
\Phi(t,z,y,\phi,\psi)= \fft{1}{z\, \sqrt{y(1-y)}}\, \Psi(t,z,y,\phi,\psi)\,,
\eea
and, noting from (\ref{tildemet}) that we have
\bea
\sqrt{-\hat g} =  \fft{8\sqrt2\, \Omega^5}{
   \sqrt{M}\, (1+ z)^2}\,,\label{tildedet}
\eea
it can be seen that if we choose $\Omega$ so that
\bea
\Omega^3 = \fft{(1+z)^2\, \sqrt{M}}{8\sqrt2}\,,
\eea
then the transformed equation (\ref{top}) is
equivalent to the following equation for $\Phi$ in the 
$\hat g_{\mu\nu}$ metric:
\bea
\del_\mu \big( \sqrt{-\hat g}\, \hat g^{\mu\nu}\,\del_\nu\Phi\big) = 0\,.
\eea
This can be derived from the Lagrangian
\bea
{\cal L} = -\ft12\sqrt{-\hat g}\, \hat g^{\mu\nu}\, \del_\mu\bar\Phi\,
\del_\nu\Phi\,.
\eea

   From the resulting energy-momentum tensor  
\bea
T_{\mu\nu}= \del_{(\mu}\bar\Phi\,\del_{\nu)}\Phi -\ft12 \hat g_{\mu\nu}\, 
\hat g^{\rho\sigma}\, \del_\rho\bar\Phi\, \del_\sigma\Phi
\eea
we may construct a conserved current $J^\mu=-K^\nu\, T^\mu{}_\nu$, where
$K=\fft{\del}{\del t}$ is the time-translation Killing vector.  This 
gives rise to a conserved energy
\bea
{\cal E} = \int \sqrt{-\hat g}\, J^0\, d^4x\,,\label{energy}
\eea
with
\bea
J^0 &=& -\hat g^{t\rho}\, \del_{(\rho}\bar\Phi\, \del_{t)}\Phi + 
   \ft12 \hat g^{\rho\sigma}\, \del_\rho\bar \Phi\, \del_\sigma\Phi\nn\\
  &=& -\ft12 \hat g^{tt}\, |\del_t\Phi|^2 + 
   \ft12 \hat g^{zz}\, |\del_z\Phi|^2 +
  \ft12 \hat g^{yy}\, |\del_y\Phi|^2 + 
   \ft12 \hat g^{\phi\phi}\, |\del_\phi\Phi|^2 +
   \ft12 \hat g^{\psi\psi}\, |\del_\psi\Phi|^2\,.\quad{}
\eea
The integrand in the energy integral (\ref{energy}) is therefore given by
\bea
\sqrt{-\hat g}\, J^0 &=& \fft1{2z^2}\, \Big\{ P\, |\del_t\Phi|^2
  + z^2\, |\del_z\Phi|^2 + y(1-y)\, |\del_y\Phi|^2 \nn\\
&&
  \qquad\qquad\qquad 
+\fft1{4(1-y)}\, |\del_\phi\Phi|^2 +
   \fft1{4y}\, |\del_\psi\Phi|^2 \Big\}\,,\label{J0}
\eea
where
\bea
P = \fft{\ep_+^2\, M}{8z^2} + \fft{\ep_-^2\,M\,  z^2}{8} +
  \fft{M}{2 z} + \fft{M\,z}{2}  +
  \ft34 M - \ft18(a^2-b^2)(1-2y)\,.\label{Pres}
\eea

The four-dimensional integration in eqn (\ref{energy}) is over the
coordinates of the 3-sphere (with the ranges $0\le y\le1$,\ \ $0\le\phi<2\pi$,
\ $0\le \psi <2\pi$), and over the transformed radial variable $z$.
This ranges over $0\le z\le \infty$, and as we discussed in section 2.4
the transformed radial function $h(z)$ for any putative unstable mode
goes rapidly to zero at $z=0$ (see eqn (\ref{hum})), and it goes
rapidly to zero at $z=\infty$ (see eqn (\ref{huminf})), ensuring the 
convergence of the integrals of all the terms in eqn (\ref{energy}).

 From eqns (\ref{eppepm2}) we have $|a^2-b^2| \le 2M$, and since 
$y$ lies in the interval $0\le y\le1$, it follows that the quantity
$P$ satisfies $P\ge0$.  Since every term in the energy integral (\ref{energy})
is integrable for any putative unstable mode, 
and each contribution is non-negative, it follows in particular
that the integral of the $\fft1{2 z^2}\,P\, |\del_t\Phi|^2$ term is bounded
from above by the conserved energy ${\cal E}$.  Thus $\Phi$ cannot grow
exponentially in time, and therefore there cannot in fact exist any 
exponentially unstable modes.

\section{3-Charge Five-Dimensional STU Supergravity Black Holes}

 The 3-charge rotating black-hole solution in five-dimensional
STU supergravity was obtained in \cite{cvetyoum5}, 
by using a 
solution-generating procedure.  A convenient form 
of the solution was given in \cite{chcvlupo}.  
With minor change of notation, to achieve consistency with our 
present conventions, the metric is given by
\bea
ds^2= (H_1 H_2 H_3)^{1/3}\, (\tx+\ty)\, d\hat s^2\,,
\eea
where
\bea
d\hat s^2= -\Phi (dt+\cA)^2 + ds_4^2\,,
\eea
with
\bea
ds_4^2 = \fft{d\tx^2}{4X} + \fft{d\ty^2}{4Y} + 
    \fft{U}{G}\, \Big(d\chi - \fft{Z}{U}\, d\sigma\Big)^2 + 
    \fft{XY}{U}\, d\sigma^2\,.
\eea
The various functions above are given by\footnote{There was one typo in
\cite{chcvlupo}: a missing factor of $(\tx+\ty)$ in the first of the
two terms in the expression for the 1-form $\cA$.  This is corrected here.}
\bea
X&=& (\tx+a^2)(\tx+b^2) -2M\, \tx\,,\qquad Y = -(a^2-\ty)(b^2-\ty)\,,\nn\\
G&=& (\tx+\ty)(\tx+\ty-2M)\,,\qquad U= \ty\, X -\tx\, Y\,,\qquad
Z=a b\, (X+Y)\,,\nn\\
\cA&=& \fft{2M c_1 c_2 c_3\, (\tx+\ty)}{G}\, [(a^2+b^2-\ty)d\sigma-
 ab d\chi] -\fft{2M s_1 s_2 s_3}{\tx+\ty}\, (ab d\sigma-\ty d\chi)\,,\nn\\
\Phi &=& \fft{G}{(\tx+\ty)^3\, H_1 H_2 H_3}\,,\qquad
  H_i = 1 + \fft{2M s_i^2}{\tx+\ty}\,.\qquad i=1,2,3.
\eea
Here $s_i=\sinh\delta_i$ and $c_i=\cosh\delta_i$, where $\delta_i$ are
the boost parameters that correspond to turning on the three electric 
charges.  When $\delta_i=0$, the metric reduces to the five-dimensional 
Myers-Perry black hole.

 The coordinates $\sigma$ and $\chi$ are related to the standard azimuthal
angular coordinates $\phi$ and $\psi$ (each with period $2\pi$) by
\bea
\sigma = \fft{a \phi - b \psi}{a^2-b^2}\,,\qquad 
\chi= \fft{b \phi -a \psi}{a^2-b^2}\,,
\eea
as can be seen from eqn (15) in \cite{chcvlupo} after turning off the gauge
coupling constant $g$.
The standard radial
and angular coordinates $r$ and $\theta$ are related to $\tx$ and $\ty$ by
\bea
\tx= r^2\,,\qquad \ty= a^2\, \cos^2\theta + b^2\, \sin^2\theta\,.
\eea
Thus $\tx$ here is the same as $\tx$ in eqn (\ref{DD}) of the uncharged case.
The coordinate $\ty$ is related to our coordinate $y=\cos^2\theta$ by
\bea
\ty = (a^2-b^2) y + b^2\,.
\eea

  Proceeding as in the earlier uncharged case, we may separate variables
and write the radial equation in the same form as eqn (\ref{Xeqn}), 
and the angular equation in the same form as 
eqn (\ref{Yeqn}).  Only the expressions for the various
$\kappa$, $\Lambda$, $\beta$ and $\gamma$ coefficients will change when the
charges are turned on.

The coefficients $\kappa$, $\Lambda$, $\beta$ and $\gamma$ in the potential 
for the radial equation, generalising those in (\ref{greek1}) for Myers-Perry,
are now given by
\bea
\kappa &=&\ft14 M \ep_+\,\ep_-\, \omega^2\,,\nn\\
\beta &=& \fft{\im}{4\sqrt{2M}}\,
 \Big[\fft{2(\Pi_c+\Pi_s)M\omega -(a+b)(m+n)}{\ep_+} -
     \fft{2(\Pi_c-\Pi_s)M\omega -(a-b)(m-n)}{\ep_-}\Big]\,,\nn\\
\gamma &=& \fft{\im}{4\sqrt{2M}}\,
 \Big[\fft{2(\Pi_c+\Pi_s) M\omega -(a+b)(m+n)}{\ep_+} +
     \fft{2(\Pi_c-\Pi_s)M\omega -(a-b)(m-n)}{\ep_-}\Big]\,,
\label{greekcharged}\\
\Lambda &=& \sigma+ \fft12 - \fft14 M(3+2s_1^2+2 s_2^2+2 s_3^2) 
 \omega^2 + \ft18 (a^2-b^2)\omega^2 \nn\\
&&
+
 \fft{[2(\Pi_c+\Pi_s)M\omega-(a+b)(m+n)]^2}{16M\ep_+^2} 
+
  \fft{[2(\Pi_c-\Pi_s) M\omega-(a-b)(m-n)]^2}{16M\ep_-^2}\,.\nn
\eea
Crucial properties that held previously in the uncharged case continue to
hold here.  In particular, since
\bea
\Pi_c+\Pi_s \ge \Pi_c-\Pi_s \ge 1\,,
\eea
together with the usual inequalities $\ep_+\le \ep_- \le1$, it follows that
when $\omega$ has a positive imaginary part, the real parts of
$\beta$ and $\gamma$ will be negative.

  For the angular equation, the hatted quantities $\hat\kappa$, $\hat\Lambda$, 
$\hat\beta$ and $\hat \gamma$ are given by
\bea
\hat\kappa &=& -\fft18 (a^2-b^2)\omega^2\,,\qquad \hat\beta=\fft{n}{2}\,,\qquad
\hat\gamma= \fft{m}{2}\nn\\
\hat\Lambda &=& \fft12 +\sigma - \fft{(m^2+n^2)}{4} +
  \fft{(a^2-b^2)\omega^2}{8}\,,\label{hattedgreeks2}
\eea
unchanged from the results (\ref{hattedgreeks}) for the uncharged
black holes.

  Following the same steps as we did previously for the uncharged Myers-Perry
black hole, we find that after implementing the same integral transformation
of the radial equation as before, we again arrive at an 
``unseparated'' equation
of the form (\ref{comb}), with the only difference being in the 
coefficient of $\omega^2$:
\bea
&&\Big\{z^2\, \del^2_{z} + y(1-y)\,\del^2_{y} -
\Big(\fft{(a+b)}{4 z} +
    \fft{(a-b)\, z}{4}\Big)\, m\omega
-\Big(\fft{(a+b)}{4 z} -
   \fft{(a-b)\, z}{4}\Big)\, n\omega\nn\\
&& +   \Big( \fft{\ep_+^2\,M}{8z^2} + \fft{\ep_-^2\, M\,  z^2}{8} +
  \fft{(\Pi_c+\Pi_s)M}{2 z} + 
  \fft{(\Pi_c-\Pi_s)M\,z}{2}  +
  \ft14 M(3+2\sum_i s_i^2)\nn\\
&&- \ft18(a^2-b^2)(1-2y)\Big)\,\omega^2
-\fft{m^2}{4(1-y)} -\fft{n^2}{4 y} + \fft1{4y(1-y)}\Big\} \Psi=0\,.
\label{3chcomb}
\eea
This correspondingly implies that the components of the tilded inverse
metric $\tilde g^{\mu\nu}$ are unchanged except for $\tilde g^{tt}$, 
which becomes
\bea
\tilde g^{tt}&=& - \Big( \fft{\ep_+^2\,M}{8z^2} + 
   \fft{\ep_-^2\, M\,  z^2}{8} +
  \fft{(\Pi_c+\Pi_s)M}{2 z} + 
   \fft{(\Pi_c-\Pi_s)M\,z}{2}  +
  \ft14 M\big(3+2\sum_i s_i^2\big) \nn\\
&&\quad\quad - \ft18(a^2-b^2)(1-2y)\Big)\,.
\eea
Calculating the determinant, we now find that instead of eqn
(\ref{tildedet}) we have
\bea
\sqrt{-\hat g} &=&  \fft{8\sqrt2\, e^{\delta_1+\delta_2+\delta_3}\,\Omega^5}{
   \sqrt{M}}\Big[(e^{\delta_1+\delta_2} + e^{\delta_3} z)
(e^{\delta_2+\delta_3} + e^{\delta_1} z)
(e^{\delta_1+\delta_3} + e^{\delta_2} z)
\nn\\
&&\qquad\qquad\qquad\qquad
(1+ e^{\delta_1+\delta_2+\delta_3}\, z)\Big]^{-\fft12}\,\,,
\label{3chtildedet}
\eea
Following the remaining steps of the previous discussion for the
uncharged case, we find that the conserved energy is given by
integrating $\sqrt{-\hat g}\, J^0$ as in equation (\ref{J0}), with
the function $P$ now given not by eqn (\ref{Pres}) but instead
\bea
P &=& \fft{\ep_+^2\,M}{8z^2} + \fft{\ep_-^2\, M\,  z^2}{8} +
  \fft{(\Pi_c+\Pi_s)M}{2 z} + 
  \fft{(\Pi_c-\Pi_s)M\,z}{2}  +
  \ft14 M(3+\sum_i 2 s_i^2) \nn\\
&&
- \ft18(a^2-b^2)(1-2y)\,.\label{3chPres}
\eea
The same arguments that established that $P$ was non-negative in the 
uncharged case show that $P\ge0$ here also, and hence again 
there cannot exist 
any unstable exponentially-growing modes.

\section{Discussion}

It has become apparent that many of the equations governing massless fields on black hole spacetimes are of Heun type (in cases where the cosmological constant is non-zero), or one of its many confluent variants, such as for non-extreme Kerr \cite{whiting}, or the extreme case \cite{rita}.  The differential equations we find for massless scalar fields in the five-dimensional Myers-Perry black hole spacetime (see \eqref{Xeqn} and \eqref{Yeqn}) are of yet another confluent Heun type.  That observation has allowed us to extend to this (and the related STU) case 
the analysis originally applied to massless fields of all spin in the Kerr spacetime \cite{whiting}.  Before our present work, that analysis had also been extended: i) by using a different integration contour, to rule out unstable modes on the real axis for the Kerr spacetime \cite{larsetal}, ii) by considering a modified integral transform, to deal with the extreme ($|a|=M$) Kerr black hole \cite{rita}, and iii), by looking carefully at more complicated examples, to establish the absence of unstable modes for massless scalar fields in STU spacetimes and all more specialized sub-cases \cite{cvgipowh}.  Remarkably, the integral transform we have used here is, effectively, an inverse of that developed for the extreme Kerr spacetime \cite{rita}. In this context, it is also worth noting that quite different techniques, stemming from Seiberg-Witten theory (see, for example \cite{Ito:2017iba}), and based on the spectral properties of the operators involved, have been used to discuss both Kerr quasi-normal modes\cite{Aminov:2020yma} and Kerr-de Sitter stability \cite{Casals:2021ugr}.  The relevance of 
such an approach to the spacetimes we consider here is yet to be determined.

\section{Conclusion}

We have shown that a massless scalar field has no exponentially unstable modes in the five-dimensional Myers-Perry black hole spacetime. We have also shown that the same is true in the five-dimensional supergravity-motivated STU spacetimes and, previously \cite{cvgipowh}, that this holds, too, for the four-dimensional STU spacetimes.  Together, these encompass a number of other special cases which arise from restricting the parameters in these more general examples.  Although these results may serve as suggestive for the behaviour for fields of higher spin -- in particular, Maxwell fields and gravitational perturbations -- it would be useful to have some more direct indication, perhaps by writing down (at least) the analogue of the Teukolsky equation in these more general cases.  That task currently remains for future work.

\section*{Acknowledgements}

We are grateful to Mihalis Dafermos, Harvey Reall and
Jorge Santos for helpful discussions.  The authors acknowledge support from the Mitchell Institute, Texas A\&M University, during the early stages of this work.  M.C.~is supported in part by DOE 
Grant Award de-sc0013528 and the Fay R. and Eugene L. Langberg Endowed Chair.
C.N.P.~is supported in part by DOE grant DE-FG02-13ER42020.  B.F.W.~is supported in part by NSF grant PHY 1607323, and thanks the Newton Institute, Cambridge, and the Institut d'Astrophysique de Paris, for hospitality while this manuscript was being written.

\appendix

\section{Global Structure of the Myers-Perry Black Hole}

   Defining a coordinate $r^*$ by the relation
\bea
dr_* = \fft{(r^2+a^2)(r^2+b^2)\, dr}{r^2\, \Delta}\,,\label{r*def}
\eea
we may introduce retarded coordinates $(u,\phi_-,\psi_-)$, where
\bea
du= dt - dr_*\,,\qquad
 d\phi_- = d\phi -\fft{a\, (r^2+b^2)\, dr}{r^2\, \Delta}\,,\qquad
d\psi_- =d\psi - \fft{b\, (r^2+a^2)\, dr}{r^2\, \Delta}\,.
\eea
In terms of these, the metric (\ref{d5mp}) then becomes
\crampest
\bea
ds^2 &=& -du^2 - 
2 dr\, (du - a \sin^2\theta\, d\phi_- -b\cos^2\theta\, d\psi_-) 
  +\rho^2\, d\theta^2 \\
&&+ 
 \fft{2M}{\rho^2}\,(du - a \sin^2\theta\, d\phi_- -b\cos^2\theta\, 
d\psi_-)^2 + (r^2+a^2)\, \sin^2\theta\, d\phi_-^2 +
(r^2+b^2)\, \cos^2\theta\, d\psi_-^2\,.\nn
\eea
\uncramp
This form of the metric is regular in the neighbourhood of
future null infinity.

   We may also introduce advanced coordinates $(v,\phi_+,\psi_+)$ by
\bea
dv=dt + dr_*\,,\qquad
 d\phi_+ = d\phi +\fft{a\, (r^2+b^2)\, dr}{r^2\, \Delta}\,,\qquad
d\psi_+ =d\psi + \fft{b\, (r^2+a^2)\, dr}{r^2\, \Delta}\,,
\eea
with respect to which the metric (\ref{d5mp}) becomes
\crampest
\bea
ds^2 &=& -dv^2 +
2 dr\, (dv - a \sin^2\theta\, d\phi_+ -b\cos^2\theta\, d\psi_+) +\rho^2\,       d\theta^2 \\
&&+
 \fft{2M}{\rho^2}\,(dv - a \sin^2\theta\, d\phi_+  -b\cos^2\theta\,
d\psi_+)^2 + (r^2+a^2)\, \sin^2\theta\, d\phi_+^2 +
(r^2+b^2)\, \cos^2\theta\, d\psi_+^2\,.\nn
\eea
\uncramp
It can be seen from this form of the metric that it is
regular as one crosses the future horizon.


\begin{thebibliography}{99}

\bibitem{whiting} B.F.~Whiting,
{\it Mode stability of the Kerr black hole},
J. Math. Phys. \textbf{30}, 1301 (1989)
doi:10.1063/1.528308

\bibitem{dafermosg} M.~Dafermos, G.~Holzegel, I.~Rodnianski and M.~Taylor,
{\it The non-linear stability of the Schwarzschild family of black holes},
[arXiv:2104.08222 [gr-qc]].

\bibitem{shlapcost} Y.~Shlapentokh-Rothman and R.~Teixeira da Costa,
{\it Boundedness and decay for the Teukolsky equation on Kerr in the full
 subextremal range $|a|<M$: frequency space analysis}, 
[arXiv:2007.07211 [gr-qc]].

\bibitem{rita}
R.~Teixeira da Costa,
{\it Mode stability for the Teukolsky equation on extremal and subextremal Kerr spacetimes},
Commun. Math. Phys. \textbf{378} (2020) no.1, 705-781
doi:10.1007/s00220-020-03796-z
[arXiv:1910.02854 [gr-qc]].

\bibitem{Shlapentokh-Rothman:2013hza}
Y.~Shlapentokh-Rothman,
{\it Quantitative Mode Stability for the Wave Equation on the Kerr Spacetime},
Annales Henri Poincare \textbf{16} (2015), 289-345
doi:10.1007/s00023-014-0315-7
[arXiv:1302.6902 [gr-qc]].

\bibitem{larsetal}
L.~Andersson, S.~Ma, C.~Paganini and B.~F.~Whiting,
{\it Mode stability on the real axis},
J. Math. Phys. \textbf{58} (2017) no.7, 072501
doi:10.1063/1.4991656
[arXiv:1607.02759 [gr-qc]].

\bibitem{cvgipowh} M.~Cveti\v c, G.W.~Gibbons, C.N.~Pope and B.F.~Whiting,
{\it Positive energy functional for massless scalars in rotating black hole 
backgrounds of maximal ungauged supergravity},
Phys. Rev. Lett. \textbf{124}, no.23, 231102 (2020)
doi:10.1103/PhysRevLett.124.231102
[arXiv:1912.08988 [gr-qc]].

\bibitem{Cvetic:1996kv}
M.~Cveti{\v c }and D.~Youm,
{\it Entropy of nonextreme charged rotating black holes in string theory},
Phys. Rev. D \textbf{54}, 2612-2620 (1996)
doi:10.1103/PhysRevD.54.2612
[arXiv:hep-th/9603147 [hep-th]]

\bibitem{Chong:2004na}
Z.~W.~Chong, M.~Cveti{\v c}, H.~Lu and C.~N.~Pope,
{\it Charged rotating black holes in four-dimensional gauged and ungauged supergravities},
Nucl. Phys. B \textbf{717}, 246-271 (2005)
doi:10.1016/j.nuclphysb.2005.03.034
[arXiv:hep-th/0411045 [hep-th]].

\bibitem{Gomez-Lobo:2009riu}
A.~G.~P.~Gomez-Lobo and J.~M.~Martin-Garcia,
{\it Spinor calculus on 5-dimensional spacetimes},
J. Math. Phys. \textbf{50} (2009), 122504
doi:10.1063/1.3256124
[arXiv:0905.2846 [gr-qc]].

\bibitem{myper} R.C.~Myers and M.J.~Perry,
{\it Black holes in higher dimensional space-times},
Annals Phys. \textbf{172}, 304 (1986)
doi:10.1016/0003-4916(86)90186-7

\bibitem{cvetyoum5} M.~Cveti\v c and D.~Youm,
{\it General rotating five-dimensional black holes of toroidally 
compactified heterotic string},
Nucl. Phys. B \textbf{476}, 118-132 (1996)
doi:10.1016/0550-3213(96)00355-0
[arXiv:hep-th/9603100 [hep-th]].

\bibitem{kerr} R.P.~Kerr,
{\it Gravitational field of a spinning mass as an example of 
algebraically special metrics},
Phys. Rev. Lett. \textbf{11}, 237-238 (1963)
doi:10.1103/PhysRevLett.11.237

\bibitem{Cvetic:1997uw}
M. Cveti{\v c} and F. Larsen,
{\it General rotating black holes in string theory: Grey body factors and event horizons,}
Phys. Rev. D \textbf{56}, 4994-5007 (1997)
doi:10.1103/PhysRevD.56.4994
[arXiv:hep-th/9705192 [hep-th]].

\bibitem{Kokkotas:1999bd}
K.~D.~Kokkotas and B.~G.~Schmidt,
{\it Quasinormal modes of stars and black holes},
Living Rev. Rel. \textbf{2} (1999), 2
doi:10.12942/lrr-1999-2
[arXiv:gr-qc/9909058 [gr-qc]].

\bibitem{chcvlupo} Z.W.~Chong, M.~Cveti\v c, H.~L\"u and C.N.~Pope,
{\it Non-extremal rotating black holes in five-dimensional gauged 
supergravity}, 
Phys. Lett. B \textbf{644}, 192-197 (2007)
doi:10.1016/j.physletb.2006.11.012
[arXiv:hep-th/0606213 [hep-th]].

\bibitem{Ito:2017iba}
K.~Ito, S.~Kanno and T.~Okubo,
{\it Quantum periods and prepotential in $ \mathcal{N}=2 $ SU(2) SQCD},
JHEP \textbf{08} (2017), 065
doi:10.1007/JHEP08(2017)065
[arXiv:1705.09120 [hep-th]].

\bibitem{Aminov:2020yma}
G.~Aminov, A.~Grassi and Y.~Hatsuda,
{\it Black Hole Quasinormal Modes and Seiberg\textendash{}Witten Theory},
Annales Henri Poincare \textbf{23} (2022) no.6, 1951-1977
doi:10.1007/s00023-021-01137-x
[arXiv:2006.06111 [hep-th]].

\bibitem{Casals:2021ugr}
M.~Casals and R.~T.~da Costa,
{\it Hidden spectral symmetries and mode stability of subextremal Kerr(-dS) black holes},
[arXiv:2105.13329 [gr-qc]].

\end{thebibliography}
\end{document}